**What is the *general Welfare*?**
**Welfare Economic Perspectives on Equity**

Charles F. Manski

Department of Economics and Institute for Policy Research
Northwestern University

October 2025

Abstract

Researchers do not know what the framers of the United States Constitution intended when they wrote of the *general Welfare*. Nevertheless, economists can conjecture by specifying social welfare functions that aim to express the preferences of the population. Economists have often simplified analysis of public policy by assuming that individuals have homogeneous, consequentialist, and self-centered preferences. In reality, individuals may hold heterogeneous private and distributional preferences. To enhance policy analysis, economists should specify social welfare functions that express the richness and variety of actual personal preferences. The possibilities are vast. I focus on preferences for *equity*. There has been much controversy regarding interpretation of equity, a term that public discourse has used in vague and conflicting ways. Specifying social welfare functions that formally express different interpretations of equity will not eliminate disagreements, but it should clarify concepts and reduce the inconsistencies that afflict verbal communication.

## 1. Introduction

A foundational objective of the Constitution of the United States is to promote the *general Welfare*. The Constitution uses the term twice. The Preamble states:

> "**We the People** of the United States, in Order to form a more perfect Union, establish Justice, insure domestic Tranquility, provide for the common defence, promote the general Welfare, and secure the Blessings of Liberty to ourselves and our Posterity, do ordain and establish this Constitution for the United States of America."

Article I, Section 8, Clause 1, sometimes called the *General Welfare Clause*, begins: "The Congress shall have Power To lay and collect Taxes, Duties, Imposts and Excises, to pay the Debts and provide for the common Defence and general Welfare of the United States."

The Constitutional premise that the United States should promote and provide for the *general Welfare* has rhetorical appeal, but it lacks substance. To make it meaningful requires a clear definition of the term. The Constitution contains no definition, nor have Acts of Congress and Supreme Court decisions clarified what constitutes the *general Welfare*. One might attempt to infer the intended meaning from parts of the Constitution such as the Bill of Rights, and from legislation that Congress has enacted and the Supreme Court has deemed constitutional. Inference might be straightforward if citizens were to hold a consensus perspective on what the nation should seek to achieve. It is evident, however, that the preferences of Americans have been heterogeneous to a considerable degree.

The vagueness of the Constitutional term *general Welfare* contrasts with the specificity of welfare economic study of public policy. Economists assume a particular social welfare function (SWF) or a well-defined class of SWFs. Research in public economics seeks to characterize the social welfare achieved by alternative feasible policies, aiming to find one that maximizes an SWF.

Economists studying policy choice in democracies strive to specify SWFs that in some manner express the values of population members rather than the preferences of dictators. The U.S. Constitution begins with the words "We the People" and refers to the *general Welfare*. It does not endorse the April 2025 proclamation of Donald Trump to journalists at the *Atlantic* that "I run the country and the world." [1]

Paul Samuelson placed responsibility for specification of the SWF on society rather than on the economist, writing (Samuelson, 1947, p. 220): "It is a legitimate exercise of economic analysis to examine the consequences of various value judgments, whether or not they are shared by the theorist." The vagueness of the term *general Welfare* and related terms has required economists to conjecture SWFs rather than draw them transparently from society.

In my own research, I have studied policy choice with what I have called *pragmatic* social welfare functions. I have characterized pragmatic SWFs as ones motivated by (Manski, 2024, p. 58): "some combination of conjecture regarding societal values, empirical study of population preferences, and concern for analytical tractability." I have found the notion of a pragmatic SWF useful. I nevertheless worry that economists commonly study SWFs that are remote from the actual heterogeneous preferences of society.[2]

Economists have mainly studied *personalist* SWFs, ones that are functions only of the personal welfare (aka *utility*) of the members of a specified population. They have, moreover, commonly presumed utilitarian aggregation of interpersonally comparable cardinal utilities.[3] My use of the term *personalist* SWF intentionally revises the word *welfarism* originated by Amartya Sen. He wrote (Sen, 1977, p. 1559): "The general approach of making no use of any information about the social states other than that of personal welfares generated in them may be called 'welfarism'." I feel that *personalist* SWF expresses Sen's intended distinction more clearly than the word *welfarism*. The *Oxford English Dictionary* defines

[1] https://www.theatlantic.com/magazine/archive/2025/06/trump-second-term-comeback/682573/, accessed June 22, 2025.

[2] Reviewing research on optimal taxation, Fleurbaey and Maniquet (2018) remarked that preference specifications often express values posited by researchers rather than the actual preferences of the population under study. They wrote (p. 1036): "Once one acknowledges that the choice of a particular utility measure is always strongly value laden, it is a small step to . . . . treat utilities as normative constructs." They then wrote that this "is probably the dominant view among optimal tax theorists."

[3] There are occasional exceptions in the literature. For example, some research on optimal tax theory has proposed aggregation of a weighted average of cardinal utilities, the weights being determined by observable personal attributes. Saez and Stantcheva (2016) wrote (p. 24): "Weights directly capture society's concerns for fairness without being necessarily tied to individual utilities." Observe their reference to "society's concerns for fairness." This phrase appears to posit an external entity, perhaps a moral philosopher, who assigns weights to members of the population.

*personalism* as follows:[4] "A system of thought according to which reality has meaning only through the conscious minds of persons, . . . . or that reality consists of interacting persons."

Sen (1977) remarked (p. 1559): "welfarism as an approach to social decisions is very restrictive." He belittled utilitarianism, writing that it presumes individuals aim to maximize "happiness," which he interpreted in a self-centered fashion. I am puzzled by Sen's perspective. Although welfare economic policy analysis has often conjectured self-centered individuals, the basic economic concept of individual utility places no restrictions on what individuals seek to maximize. In principle, some individuals may be liberal moral philosophers, holding normative beliefs that society should adherence to notions of justice, fairness, and equity. Others may want to enhance inequalities in directions they favor and may advocate discrimination against members of certain groups. Each individual may hold a distinct view on the composition of the population—past, present, and future generations of humans and other forms of life—whose interests should be considered when measuring social welfare.

All of these possibilities may be embraced within the general concept of a personalist SWF and the particular one of utilitarian planning. The possibilities are too vast to explore fully in one paper. To make the scope of discussion manageable, I consider personalist SWFs in which individuals may have distributional preferences for *equity*. There has been much controversy regarding the meaning of equity. Public discourse has used the word in vague and conflicting ways. The related words *equality* and *fairness* have also been used loosely.

To eliminate vagueness, I study SWFs that formally express different interpretations of equity. Formal analysis will not eliminate disagreements about the types of equity that individuals value. However, it should reduce the misunderstandings and logical inconsistencies that now afflict verbal communication.

In the 20th century, economists used various devices to circumvent or grossly simplify specification of SWFs. These include the Hicks (1939) introduction of the *new welfare economics* and the proposal by Kaldor (1939) and Hicks (1939) that economists use the artifice of Kaldor-Hicks efficiency to separate the study of Pareto efficiency and the distribution of personal welfare. They include the macroeconomic practice of using *representative-agent* models to avoid consideration of heterogeneity in personal preferences. Atkinson (2011) referred to these devices as "avoidance strategies" and called for "The Restoration of Welfare Economics" as a central subject of study by economists. I agree with Atkinson. This paper aims to follow up on his call.

[4] https://www.oed.com/dictionary/personalism_n?tab=meaning_and_use#30959752, accessed June 22, 2025.

Section 2 opens discussion of personal preferences for equity, aiming to clarify important distinctions that are not always appreciated. I point out that it commonly is logically impossible to jointly achieve different types of equity. I call attention to multiple interpretations of horizontal equity and equal treatment of equals. I discuss conflicting concepts of health equity in research on public health, and a controversy regarding intergenerational equity in research on climate policy. I observe that alternative perspectives on equity are embedded in different mechanisms for aggregating personal preferences into a personalist SWF.

Whereas the discussion in Section 2 is verbal, Section 3 formally specifies SWFs that express preferences for equity. I present basic principles in abstraction, but productive analysis requires concreteness. I do this in steps. The first is to consider policies that give each member of the population a choice set, from which the member chooses an action. The second is to assume that individual utility has a self-centered component and a distributional component, the latter expressing personal preferences for equity. The third is to consider utilitarian social welfare when individual utility is additively separable in self-centered and distributional preferences, the latter being concerned with equality of opportunity. I then illustrate optimal utilitarian policy, showing how the optimum depends on the strength and type of distributional preferences.

I report two very simple findings that I think important enough to label as propositions. The first shows that, in settings with monotonic separability of preferences and infinitesimal interactions (to be defined), rational individual choice behavior under a specified policy is self-centered regardless of distributional preferences. Yet this does not imply that an individual would, if empowered to be the social planner, choose a policy to maximize the self-centered component of utility. Acting as the planner, an individual with distributional preferences may choose to sacrifice some personal prosperity in order to enhance population equity.

The second finding shows the important of distinguishing complete equality of opportunity, where all members of the population face the same choice set, from intergroup equality, where different groups face the same distribution of choice sets. I point out that intergroup equality is consistent with any degree of intragroup variation in choice sets. For example, racial equality of opportunity is consistent with extreme inequality across individuals of each race.

## 2. Preferences for Equity

In principle, a personalist SWF can aggregate the welfare of individuals who hold heterogeneous consequentialist and deontological preferences over abstract social states. Arrow (1978) put it this way (p. 224):[5]

> What remains is the determination of the social ordering. On what data is it based? In particular, how does it relate to individual preferences over social states, what might be termed, "individual utilities." For purpose of this paper, I am accepting the viewpoint of the utilitarians and of welfare economics. It is assumed that each individual has some measure of the satisfaction he draws from each social state and that the social ordering is determined by the specification of these utilities for all possible social states."

In practice, economists rarely consider personal preferences over abstract social states. The convention has been to assume that preferences are consequentialist and self-centered, sometimes called egoism (Driver, 2022). Fleurbaey (2021) stated this in striking fashion, writing (p. 39):

> "It is standard in normative economics, as in political philosophy, to evaluate individual well-being on the basis of self-centered preferences, utility or advantage. Feelings of altruism, jealousy, etc. are ignored in order not to make the allocation of resources depend on the contingent distribution of benevolent and malevolent feelings among the population."

[5] Arrow's 1978 statement differs notably from his argument against utilitarian aggregation of utilities in his famous 1951 book. He wrote there that it (Arrow, 1951, p. 11): "seems to make no sense to add the utility of one individual, a psychic magnitude in his mind, with the utility of another individual."

Fleurbaey did not argue that it is realistic to assume that personal preferences are self-centered. He only remarked that this assumption simplifies analysis.

A vast array of possible preferences over social states opens when one entertains the possibility that individuals need not have purely consequentialist and self-centered preferences. Experimental economists has accumulated empirical evidence for various forms of *other-regarding* (or social) preferences; see Cooper and Kagel (2017) and Fehr and Charness (2025). Concluding their review article, Fehr and Charness call for recognition of social preferences in welfare economics, writing (p. 503):

> "it would be desirable to study the deeper implications of heterogenous social preferences for normative (public) economics. If individuals display altruistic or inequality averse distributional preferences, it does not make much sense to compute optimal policies on the basis of social welfare functions that assume that every individual only cares for his or her own consumption. Isn't economics, after all, built on a deep commitment to respect individuals' preferences? Likewise, if people care also for equality of opportunity, it appears of paramount importance to incorporate that notion into modern welfare economics rather than computing optimal policies on the basis of a standard utilitarian welfare function that assumes that individuals only care for their own consumption."

Social preferences need not favor equity—some individuals may want to increase some inequalities. I will, however, focus on preference for equity, which has been a subject of intense public discussion and controversy. The controversy persists in part because the word *equity* has been used in distinct and often vague ways. I aim to clarify the distinctions and sharpen the discourse.

Section 2.1 observes that it is often impossible to jointly achieve different types of equity. Sections 2.2 and 2.3 discuss various interpretations of the ideas of horizontal equity and equal treatment of equals. Sections 2.4 and 2.5 focus on two substantive contexts, health equity and intergenerational equity. Section 2.6 remarks that alternative perspectives on equity are embedded in different mechanisms for aggregating personal preferences into SWFs.

## 2.1. The Impossibility of Jointly Achieving Different Types of Equity

To begin, I call attention to the important but underappreciated fact that it is often logically impossible to jointly achieve different types of equity.

A simple example occurs in medicine. Consider groups of persons who vary in their response to a given treatment of an illness, the treatment being more effective in curing some groups than others. Equalizing the rate at which these groups receive the treatment will yield disparities in their health outcomes. Equalizing health outcomes across groups will require disparities in the rate at which they receive the treatment. It is thus impossible to equalize both rates of treatment and health outcomes across groups. See Section 2.4 for further discussion.

In what appears to be the first formal analysis of this logical conundrum, Dominitz (2003) called attention to the impossibility of achieving different types of equity in a controversial aspect of criminal justice policy: racial profiling in traffic stops searching for illegal drugs. Racial profiling has been criticized as a deleterious inequity. Dominitz observed that profiling has been measured in multiple ways: by comparison of race-specific search rates, drug find rates, thoroughness of search, rates of detention of the innocent, and rates of apprehension of the guilty. He showed that, if crime rates differ across races, it is logically impossible to simultaneously eliminate disparities in all of these respects. He cautioned (p. 415): "policy makers must decide whether to sacrifice equality of detention rates of the innocent, equality of apprehension rates of the guilty, or both, because they cannot be simultaneously satisfied." [6]

[6] Twenty years later, Green (2022) labelled this type of finding 'the impossibility of fairness,' citing research on algorithmic fairness by Chouldechova (2017) and Kleinberg et al. (2016).

## 2.2. Varieties of Horizontal Equity

In their textbook on public economics, Atkinson and Stiglitz (1980) distinguished several broad ways in which economists have used the term *horizontal equity*. They wrote (p. 293): "The principle of horizontal equity states that those who are in all relevant senses identical should be treated identically."[7] They juxtaposed three interpretations of horizontal equity that they had observed in the public economics literature:

(1) (p. 294): "horizontal equity is simply an implication of the more general principle of welfare maximization."

(2) (p. 294): "it is an independent principle of justice, which has to be set into the balance alongside maximization of welfare."

(3) (p. 295): "a restriction on instruments rather than as based on a comparison of distributions."

In interpretation (1), Atkinson and Stiglitz meant that analysis of policy choice assuming certain personalist SWFs and certain policy spaces yields the conclusion that homogeneous treatment of observationally identical persons maximizes social welfare. Interpretation (2) makes horizontal equity a non-personalist ethical consideration, which society may want to consider along with personalist welfare. Interpretation (3) views horizontal equity lexicographically, as an ethical mandate constraining the space of policies that society views as feasible.

Atkinson and Stiglitz emphasized that the interpretations differ, writing (p. 295): "These three interpretations of horizontal equity are rather different. The first and second are concerned with the *results* of policy; the third is concerned with the *means* used to achieve the results."

[7] Considering people who are not identical in all relevant aspects, Atkinson and Stiglitz (1980) also observed that public economists studying taxation have referred to various forms of *vertical equity* and the related term *ability to pay*, advocating principles of *equal sacrifice* or *equal marginal sacrifice*. See Sections 11-4 and 13-1.

## 2.3. Ex Ante and Ex Post Equal Treatment of Equals

The phrase *equal treatment of equals* is sometimes used as a synonym for horizontal equity. When studying policy choice under uncertainty, it is important to distinguish between ex ante and ex post equal treatment of equals.

In Manski (2009, 2024) and elsewhere, I have studied treatment diversification as an approach to coping with uncertainty regarding policy outcomes. A simple and instructive case is utilitarian allocation of a population of observationally identical individuals to two feasible treatments. Contemplating assignment of everyone to a single treatment, suppose that a planner does not know which treatment yields larger social welfare, but the planner can bound these welfare values. In this setting, I have shown that a specific fractional (aka diversified) allocation minimizes maximum regret.

I have called attention to the fact that diversification yields equal treatment of equals in the *ex-ante* sense that all members of the population have the same probability of receiving a particular treatment. Yet it violates equal treatment in the *ex-post* sense that different persons ultimately receive different treatments. Thus, equal treatment holds ex ante but not ex post.[8]

Atkinson and Stiglitz (1980) recognized the distinction between ex-ante and ex-post equal treatment. They wrote (p. 296): "In some writing on welfare economics, it has been assumed that *ex ante* welfare is the natural welfare function; others would argue that the *ex-ante* criterion is unacceptable and even, indeed, unconstitutional as a basis for taxation." They did not write more deeply on the subject because they focused on policy choice in the absence of uncertainty about policy impacts. In contrast, uncertainty has been a central concern in my research on policy choice.

Manski (2009) observed that democratic societies usually adhere to the ex post sense of equal treatment. Americans with identical income, deductions, and exemptions are required to pay the same federal income tax. The Equal Protection clause in the 14th Amendment to the Constitution is held to mean that all persons in a jurisdiction are subject to the same laws, not that all persons have the same chance of being subject to different laws.

[8] The distinction between ex ante and ex post equal treatment in Manski (2009, 2024) should not be confused with the conceptually different one made in Fleurbaey and Peragine (2013). Whereas the former distinction concerns resolution of uncertainty, the latter concerns resolution of effort decisions made by individuals.

Nevertheless, some accepted policies yield equal treatment ex ante but not ex post. American examples include random tax audits, drug testing and airport screening, calls for jury service, and the Green Card and Vietnam draft lotteries. These policies have not been prompted by the desire to cope with uncertainty that motivates treatment diversification. Yet they indicate some willingness of society to accept policies that provide equal treatment ex ante but not ex post.

Acceptance of equal treatment ex ante but not ex post is especially notable in performance of randomized experiments, undertaken to learn about treatment response. Combining ex ante equal treatment with ex post unequal treatment is precisely what makes randomized experiments informative. Modern medical ethics permits randomization only under conditions of *clinical equipoise*; that is, when partial knowledge of treatment response prevents an ex ante determination that one treatment is superior to another.

## 2.4. Varying Perspectives on Health Equity

The term *health equity* has become familiar, with extraordinary breadth and vagueness of usage. Braverman (2006) describes the modern history of the term in public health research. The abstract to her article begins with the statement (p. 167): "There is little consensus about the meaning of the terms "'health disparities," "health inequalities," or "health equity." Among recent definitions by public health agencies, the U.S. Centers for Disease Control and Prevention states: "Health equity is the state in which everyone has a fair and just opportunity to attain their highest level of health."[9] The U.S. National Cancer Institute states: "A situation in which all people are given the chance to live as healthy a life as possible regardless of their race, ethnicity, sex, sexual orientation, disability, education, job, religion, language, where they live, or other factors."[10] The World Health Organization states: "Health is a fundamental human right. Health equity is achieved when everyone can attain their full potential for health and well-being."[11]

[9] https://www.cdc.gov/health-equity/what-is/index.html. Accessed May 24, 2025.

Economist have made diverse contributions to the conceptualization and measurement of health equity. Fleurbaey and Schokkaert (2011) review a considerable literature spanning economics and philosophy. I focus here on the utilitarian perspective, which sits fully within classical welfare economics.

### 2.4.1. The Utilitarian Perspective

Health economists have viewed clinicians as utilitarian planners who treat populations of patients. A clinician observes certain covariates for each patient, who has some risk of illness. The objective is to maximize mean health-related utility. It is common to assume that care is individualistic, meaning that the care received by one patient may affect that person but does not affect others. This assumption is generally realistic when considering non-infectious diseases.

A common problem in clinical decision making is that treatments must be chosen with incomplete knowledge of their health outcomes. Health economists often assume that the clinician knows the objective probability distribution of personal outcomes that will occur if a patient with specified observed covariates is given a specified treatment; that is, the clinician has rational expectations. In this setting, the problem of optimizing utilitarian patient care has a simple solution: patients should be divided into groups having the same observed covariates and all patients in a group should be given the care that yields the highest within-group mean patient welfare. Patients with the same observed covariates should be treated uniformly.

Achievable utilitarian welfare weakly increases as more patient covariates are observed. Observing more covariates enables a clinician to refine the probabilistic predictions of treatment outcomes on which decisions are based. Refining these predictions is beneficial if doing so affects optimal treatment choices. This result has been discussed by Phelps and Mushlin (1988), Basu and Meltzer (2007), Manski (2013), and elsewhere. Manski, Mullahy, and Venkataramani (2023) provide a proof in the simple setting of choice between two treatments.

---

[10] https://www.cancer.gov/publications/dictionaries/cancer-terms/def/health-equity. Accessed May 24, 2025.

[11] https://www.who.int/health-topics/health-equity#tab=tab_1. Accessed May 24, 2025.

Utilitarian treatment choice with rational expectations formally implements a perspective on the idea that medical decision making should be equitable: clinicians should do as well as possible for their patients, on average given what is known about them. Utilitarian optimization does not imply that patients with different observed covariates should receive the same treatment or that they will experience the same health. The utilitarian sense of equity may be achieved with differences in treatments and health outcomes across groups of patients. Recall the discussion in Section 2.1, pointing out that it is impossible to equalize both rates of treatment and health outcomes across groups if groups of patients vary in their response to treatment of an illness.

2.4.2. Non-Utilitarian Perspectives on the Use of Measures of Race to Make Clinical Predictions

A growing segment of the medical community in the United States have deemed certain treatment and health disparities undesirable from non-utilitarian perspectives on health equity, particularly when the disparities are by race. An influential movement to remove race as a covariate in existing algorithms for medical risk prediction has developed.

Commentaries by Cerdeña *et al.* (2020), Vyas *et al.* (2020), and Briggs (2022) exemplify calls to cease the use of race as a covariate. Cerdeña *et al.* (2020) stated that (p. 1125): "race-based medicine . . . ., perpetuates health-care disparities." Vyas *et al.* (2020) called for a general reconsideration of the use of race in risk assessments. After documenting the use of race as a covariate in algorithms predicting patient outcomes in many fields of medicine, they asserted (p. 874): "Many of these race-adjusted algorithms guide decisions in ways that may direct more attention or resources to white patients than to members of racial and ethnic minorities. . . . . Given their potential to perpetuate or even amplify race-based health inequities, they merit thorough scrutiny." Briggs (2022) wrote (p. 2116):

> "while it is difficult to refute the central contention that optimal decision making requires the use of all covariates that are associated with outcome, the assumption that racial covariates, and their

> application within the medical arena, are sufficiently free from bias (structural, institutional or personal) misses the point of the underlying argument: that race is not the same as every other covariate in our arsenal. It is a covariate that is acting as a proxy for a wide range of other explanatory variables that could be genetic/biological but in many circumstances are more likely to be sociological/socioeconomic."

Leading institutions have recommended race-free risk prediction. A notable case is Delgado *et al.* (2021), which recommended removal of race as a predictor of kidney disease. This recommendation has since been implemented in major medical centers.

Manski (2022) and Manski, Mullahy, and Venkataramani (2023) questioned four assertions that have been advanced as arguments against the inclusion of race as a covariate in medical risk prediction. These assertions are: (i) race is a social, not biological, concept. (ii) there is no established causal link between race and the illness. (iii) using race may perpetuate or worsen racial health inequities. (iv) many persons are offended by the use of race in risk assessment.

We observed that assertions (i) and (ii) are not relevant to assessment of race as an informative predictor of illness. Usefulness in prediction only requires statistical association, not an understanding of the reasons for association. Advocates of removing race from medical risk assessments have not made clear the logic of claiming that use of race as a predictor in medical risk assessment is not appropriate if race is a social construct. Nor have they made clear the logic of concern with causality, given that statistical association suffices for successful prediction within a population.

We observed that assertions (i), (iii), and (iv) are empirical assertions that have been the subject of considerable controversy, with evidence being scant. It is not known to what extent patient populations would be willing to give up some accuracy in the medical predictions made for them, in order to mitigate the types of racial disparities that medical commentators have argued are undesirable. Nor has there been much empirical study of how elimination of race as a predictor affects the magnitudes of the various types of disparities that commentators have deemed problematic.

## 2.5. Preferences for Intergenerational Equity

Macroeconomists commonly use present discounted Gross Domestic or World Product as a surrogate for the *general Welfare*. In so doing, they ignore the distribution of income among persons who are currently alive. Discounting the future takes an explicit stand on the distribution of income across generations, with the discount rate expressing preferences for intergenerational equity.

It is not feasible to elicit the preferences of future humans who are as yet unborn. Elicitation of preferences for intergenerational equity is feasible in principle from the population who are alive currently, but that is not the practice.[12] Rather, the discount rate is a choice made by researchers. I use economic analysis of climate policy to illustrate.

### 2.5.1. Specification of the Discount Rate in Analysis of Climate Policy

*Integrated assessment* (IA) models enable quantitative evaluation of alternative climate policies. IA models provide long-run descriptions of the global economy, including the energy system and its role in economic production. They represent the climate and the links between the climatic effects of greenhouse gas (GHG) emissions and their impacts on the economy. IA models have become primary tools for comparing policies to reduce GHG emissions. A leading example is the Dynamic Integrated Climate Economy (DICE) model described in Nordhaus (2019).

In DICE and similar models, the economic losses from climate change are represented by damage functions that give the decreases in world-wide output resulting from increases in mean global temperature. Policy comparisons have been performed by considering a planner who seeks to make optimal trade-offs between the costs of carbon abatement and the global economic damages from climate change. The planner is assumed to face an optimal-control problem, the objective being to minimize the present discounted costs of abatement and damages over a time horizon. Thus, present discounted gross world product expresses social welfare.

---

[12] While elicitation of the preferences of the current population for intergenerational equity has not been the practice, it is a central premise of *The Ministry for the Future* (Robinson, 2020). In his speculative novel, Robinson conjectures creation of a United Nations organization whose mission is (p. 16): "to advocate for the world's future generations of citizens, whose rights, as defined in the Universal Declaration of Human Rights, are as valid as our own."

The use of discounted world product to express welfare ignores the distribution of welfare among persons who are currently alive.[13] It does not ignore intergenerational equity. The specified discount rate quantifies the present value of the future costs and benefits resulting from climate policy. The appropriate discount rate has been a contentious issue in climate economics. Controversy persists in part due to the fact that the choice of an appropriate discount rate is not only an empirical question concerning the state of the economy in the future. It is a normative matter concerning preferences for equity across future generations. Economists commonly use some version of the *Ramsey formula* (Ramsey, 1928) to expression the interplay of normative and empirical considerations in choosing a discount rate. See Dasgupta (2008).

The specified discount rate can be a highly consequential determinant of conclusions on optimal climate policy. This became plain when Nordhaus (2007) and Stern (2007) reported the findings of studies using different discount rates. Stern used a relatively low rate and concluded that policy should seek to reduce GHG emissions aggressively and rapidly. Nordhaus used a relatively high rate and concluded that policy should act moderately and slowly.

## 2.6. Perspectives on Equity in Mechanisms for Preference Aggregation

To the extent that personal preferences are heterogeneous, optimal policy choice is necessarily sensitive to the mechanism for preference aggregation used in a personalist social welfare function.

[13] Discussions of climate policy often refer to distributional impacts, but only verbally. Consider, for example, a report on decarbonization policy published by a committee of the National Academies of Science, Engineering, and Medicine (2023). The Committee wrote that society should aim to achieve (p. 4): "a fair, equitable, and just 30-year transition" to decarbonization. It went on to write (p. 51):

> "the objective to reduce emissions to net zero by midcentury (or 50 percent by 2030) can be thought of as a constraint with the goal to minimize cost while maximizing desirable societal objectives of equity, employment, health, and public engagement."

The Committee did not formalize the terms "fair, equitable, and just." Nor did it assess what society should do if the "objectives of equity, employment, health, and public engagement" should be in tension with one another.

Different aggregation mechanisms embody different perspectives on equity. Welfare economics has viewed the mechanism as a normative meta-decision made by a social planner, who stands outside of society and acts in its behalf.

### 2.6.1. Harsanyi and Rawls As-If Consensus on the Aggregation Mechanism

Seeking to overcome the need to posit an external social planner to aggregate preferences, Harsanyi (1955) and Rawls (1971) conjured a thought experiment that they interpreted in competing ways, yielding dramatically different perspectives on equity. Harsanyi and Rawls both asserted that society as a whole would have consensus in favor of a particular mechanism. However, they disagreed on what this consensus would be. For Harsanyi, it was utilitarian aggregation of interpersonally comparable cardinal utilities. For Rawls, it was maximin evaluation of interpersonally comparable ordinal utilities.[14] I begin with the Rawls argument, which is more widely known, even though the Harsanyi proposal preceded it by about fifteen years.

Rawls argued that the aggregation mechanism should be determined by a social contract. He maintained that this contract should express a consensus that he argued all rational people would accept in an *initial position*, characterized by a *veil of ignorance*. He wrote (p. 10):

> "the guiding idea is that the principles of justice for the basic structure of society are the object of the original agreement. They are the principles that free and rational persons concerned to further their own interests would accept in an initial position of equality."

He declared that he knew what principles free and rational persons would accept, writing (p. 13):

> "I shall maintain instead that the persons in the initial situation would choose two rather different principles: the first requires equality in the assignment of basic rights and duties, while the second holds that social and economic inequalities, for example inequalities of wealth and authority, are just only if they result in compensating benefits for everyone, and in particular for the least advantaged members of society."

---

[14] A distinction between the Harsanyi and Rawls arguments is that Harsanyi, but not Rawls, considered an individual's overall measurement of utility. Rawls conjectured a lexicographic process in which an individual first establishes certain principles of justice and then evaluates his material circumstances. He applied his maximin argument only to what he called *primary goods*, not the entirety of material circumstances.

Thus, Rawls assumed that personal welfare is ordinally comparable across individuals. and he argued that social welfare should be measured as the minimum personal welfare of all members of society.

Rawls did not originate the idea that all rational people would agree on the aggregation mechanism in a hypothetical original position. Harsanyi (1955) posed a thought experiment of this type and reached a different conclusion. He argued that, not knowing their positions in society, individuals in the original position would place equal probability on realizing each possible position and would maximize expected utility. He thus concluded that all rational persons would accept a utilitarian SWF.[15]

The Harsanyi and Rawls arguments differ fundamentally in their perspectives on equity. Maximization of expected utility with equal probabilities of each possible initial position yields ex ante equal treatment of equals. It provides no guarantee against extreme inequality in personal welfares ex post. Maximization of minimum utility guarantees against extreme deprivation in ex post minimum welfare. It is insensitive to the distribution of welfare above the minimum.

Critics have questioned how one could know that all free and rational persons would accept either the Harsanyi or Rawls principles. In his review of the Rawls book, Arrow (1973) wrote (p. 247):

> "How do we know other peoples' welfare enough to apply a principle of justice?" . . . ."the criterion of universalizability may be impossible to achieve when people are really different, particularly when different life experiences mean that they can never have the same information."

He concluded by writing (p. 263):

> "To the extent that individuals are really individual, each an autonomous end in himself, to that extent they must be somewhat mysterious and inaccessible to each other. There cannot be any rule that is completely acceptable to all."

[15] Writing fifteen years later, Rawls barely acknowledged the Harsanyi argument, mentioning Harsanyi by name only briefly in a footnote. Nevertheless, he attacked utilitarianism sharply.

## 3. Social Welfare Functions Expressing Personal Preferences for Equity

Specification of SWFs expressing alternative interpretations of equity should reduce the misunderstandings and logical inconsistencies that presently afflict public discourse. I now proceed.

Section 3.1 begins abstractly. Abstraction enables transparent presentation of basic principles, but productive analysis requires concreteness. As a first step towards concreteness, I consider policies that give each member of the population a choice set, from which the member chooses an action.

In a second step towards concreteness, Section 3.2 assumes that individual utility has a self-centered component and a distributional component. The distributional component expresses personal preferences for equity. I show that, in settings with monotonic separability and infinitesimal interactions (to be defined), rational individual choice behavior under a specified policy is self-centered regardless of distributional preferences. I then distinguish various properties of the distribution of circumstances, including equality of choice sets (aka equal opportunity), equality of chosen actions, and equality of realized utility. In each case, I distinguish complete and intergroup equality.

In a further step toward concreteness, Section 3.3 formalizes utilitarian social welfare when individual utility is additively separable in self-centered and distributional preferences, the distribution preference being concerned with equality of opportunity. Section 3.4 presents two illustrations of optimal utilitarian policy, one assuming that individuals prefer complete equality and the other that they prefer intergroup equality. The illustrations show explicitly how optimal policy depends on the strength and type of distributional preferences. I conclude with a brief discussion of numerical computation of optimal utilitarian policy (Section 3.5) and the need for empirical research to learn population preferences (Section 3.6).

### 3.1. Personalist Social Welfare in Abstraction

Welfare economists have occasionally posited that individuals have preferences over abstract social states. See, for example, Arrow (1951, 1978). A personalist SWF aggregates these preferences in some manner.

Let J denote the population under consideration. Let $\Phi$ denote the set of feasible policies and Z denote the set of feasible social states. For each policy $\varphi \in \Phi$, $z_\varphi \in Z$ is the social state attained under $\varphi$. Let each individual $j \in J$ have a preference ordering on Z, admitting a utility function $u_j(\cdot): Z \to R$. If individual j were empowered to be the social planner, with the capacity to choose a policy unilaterally, j would choose $\varphi$ to maximize $u_j(z_\varphi)$.

A personalist SWF is a real function $W(\cdot)$ of the utilities of all members of J. The social welfare of policy $\varphi$ is $W[u_j(z_\varphi), j \in J]$. A planner with SWF $W(\cdot)$ chooses $\varphi$ to maximize $W[u_j(z_\varphi), j \in J]$. Research in welfare economics has generally assumed that social welfare respects personal preferences, in the sense that $W(\cdot)$ is strictly monotone in each individual utility.

Individual utilities are said to be self-centered if each social state z is a vector $z = (z_j, j \in J)$ and, for each j, $u_j(z)$ varies with z only through $z_j$. Personal preferences need not have this structure. In principle, $u_j(z)$ may vary with z in any manner. A central consideration in this paper is that individual utility functions need not be self-centered.

It is important to recognize that making $W(\cdot)$ a personalist SWF does not imply that individual utility functions are themselves personalist. That is, $u_j(\cdot)$ may vary with z in ways other than through the values of $u_k(z)$, $k \in J$. The utility functions with distributional preferences introduced in Section 3.2 are not personalist.

#### 3.1.1. Policies Specifying Choice Sets

As a first step from abstraction to concreteness, I henceforth consider the class of policies that give each member of the population a choice set. Under policy $\varphi$, individual $j \in J$ faces choice set $C_{\varphi j}$ and chooses some action $c_{\varphi j} \in C_{\varphi j}$. Individuals have personal covariates $(x_k, k \in J)$, taking values in some set X, that do not vary with policy. Covariates may, for example, include demographic attributes and utility functions. The social state under policy $\varphi$ is a vector $z_\varphi = (x_j, c_{\varphi j}, C_{\varphi j}; j \in J)$.

A prominent example is income tax policy. Here $\varphi$ is a tax structure, $C_{\varphi j}$ is the set of (consumption, leisure) pairs available to j under $\varphi$, and $c_{\varphi j}$ is the choice that j would make. Another prominent example is medical treatment. Here $\varphi$ is a health policy, $C_{\varphi j}$ is the set of treatments available to j under $\varphi$, and $c_{\varphi j}$ is the treatment that j would choose, perhaps in conjunction with a clinician.

Economists studying income tax and health policy commonly assume that personal preferences are self-centered. They also assume that utility is a function only of the choice $c_{\varphi j}$ that an individual makes, not the choice set $C_{\varphi j}$. Thus, $u_j(z_\varphi) = u_j(c_{\varphi j})$. These assumptions need not hold. In principle, $u_j(\cdot)$ may be a function of the covariates, choices, and choice sets of all members of the population.

## 3.2. Distributional Preferences

Public discourse on policy typically concerns the distribution of circumstances in a population, not the circumstances of particular named individuals. For each policy $\varphi$, let $P(x, c_\varphi, C_\varphi)$ denote the population distribution of covariates, choices, and choice sets. With this notation, I now formalize *distributional preferences*. I assume that $u_j(\cdot)$ is a function of j's own choice and of the distribution $P(x, c_\varphi, C_\varphi)$. Thus,

$$(1) \qquad u_j(z_\varphi) = u_j[c_{\varphi j}, P(x, c_\varphi, C_\varphi)].$$

Observe that utility function (1) is not personalist. Individual j may have a preference over the entire population distribution of circumstances, not just the distribution of realized utility.

Among utility functions of this form, considerable simplification in analysis of individual choice behavior occurs if $u_j(\cdot)$ is monotonically separable in its self-centered component and if population interactions are infinitesimal. Monotonic separability means that

$$(2) \qquad u_j(z_\varphi) = f_j[v_j(c_{\varphi j}), P(x, c_\varphi, C_\varphi)],$$

where $v_j(\cdot)$ and $f_j(\cdot)$ are real-valued and where $f_j(\cdot)$ is strictly increasing in $v_j(\cdot)$. Infinitesimal interactions means that the action chosen by individual j does not affect the distribution $P(x, c_\varphi, C_\varphi)$. Given these conditions, we have this simple and powerful result:[16]

Proposition 1: Assume monotonic separability and infinitesimal interactions. Then rational individual choice behavior under a specified policy is self-centered regardless of distributional preferences. ■

Proof: Fix policy $\varphi$. By the assumption of infinitesimal interactions, $P(x, c_\varphi, C_\varphi)$ is invariant to the action chosen by j. Hence, utility maximization by j implies that $c_{\varphi j}$ maximizes $v_j(d)$ over $d \in C_{\varphi j}$.

Q. E. D.

Proposition 1 characterizes the behavior of individual j under a specified policy. It does not imply that j would, if empowered to be the social planner, choose a policy to maximize the self-centered component of utility. If acting as the social planner, an individual with distributional preferences may choose to sacrifice some personal prosperity in order to enhance the population distribution of circumstances.

Thus far, a utility function with distributional preferences may vary with the entire distribution $P(x, c_\varphi, C_\varphi)$. Many distinct interpretations of preferences for equity refer to particular properties of this distribution. The sub-sections below distinguish equality of opportunity, equality of chosen actions, and equality of realized individual utility.[17] In each case, I distinguish *complete equality* and *intergroup equality*.

[16] Monotonic separability and infinitesimal interactions seem realistic assumptions in many settings. However, the conclusion of Proposition 1 is at odds with some empirical observations of apparently altruistic behavior, such as when individuals make small donations to large charities.

[17] These concepts have previously been distinguished from a philosophical perspective. See, for example, Fleurbaey (1995). In contrast, my concern is with the types of equality that population members may value.

3.2.1. Equality of Opportunity

Complete equality of opportunity under policy $\varphi$ means that almost all members of the population face the same choice set. Formally, $P(C_\varphi)$ is degenerate. Complete equality of opportunity obviously is a very strong property.

Much recent public discourse has concerned weaker concepts of intergroup equality, for example racial or gender equality. To formalize these ideas, use the covariates x to classify the population into a set G of mutually exclusive and exhaustive groups ($X_g \subset X$, $g \in G$). Let $P(C_\varphi|x \in X_g)$ denote the distribution of choice sets for persons with covariates in $X_g$. Intergroup equality of opportunity under policy $\varphi$ means that every group faces the same distribution of choice sets. That is, $P(C_\varphi|x \in X_g) = P(C_\varphi|x \in X_{g'})$ for all distinct groups g and g′.

It is important not to confuse complete and intergroup equality. The distinction is so important that I label it as a proposition, whose validity is immediate:

<u>Proposition 2</u>: Intergroup equality of opportunity is consistent with any degree of intragroup variation in choice sets.[18] ■

For example, racial equality of opportunity does not imply that all members of a given racial group face the same choice set. Racial equality is consistent with extreme inequality across individuals of each race.

3.2.2. Equality of Chosen Actions

Complete equality of chosen actions under policy $\varphi$ means that almost all members of the population make the same choice; thus, $P(c_\varphi)$ is degenerate. If individuals with heterogeneous preferences all face the same non-singleton choice set, they may make different choices. As illustrated in the medical treatment example of Section 2, it is not generally possible to achieve complete equality of both opportunity and chosen actions if individual preferences are heterogeneous and policy $\varphi$ does not mandate one action.

[18] Intergroup equality is equivalent to complete equality only in the limit case where G = J. Then each $g \in G$ names one individual. Hence, intergroup equality means that $P(C_\varphi|x \in X_g)$ is degenerate at the same choice set for all persons.

To formalize intergroup equality, again classify the population into groups $(X_g, g \in G)$. Let $P(c_\varphi|x \in X_g)$ denote the distribution of chosen actions for persons with covariates in $X_g$. Intergroup equality of chosen actions under policy $\varphi$ means that $P(c_\varphi|x \in X_g) = P(c_\varphi|x \in X_{g'})$ for all groups g and g′. As in Proposition 2, intergroup equality of chosen actions is consistent with any degree of intragroup variation in chosen actions.

### 3.2.3. Equality of Realized Utility

Choice sets and chosen actions are objective quantities that may be observable. Individual utilities are subjective and generally are not observable. Nevertheless, public discourse sometimes refers to equality of personal welfare. This concept is well-defined if utilities are interpersonally comparable. Then we may interpret complete equality of realized utility under policy $\varphi$ to mean that $P[u(z_\varphi)]$ is degenerate. Intergroup equality means that $P[u(z_\varphi)|x \in X_g] = P[u(z_\varphi)|x \in X_{g'}]$ for all groups g and g′.

## 3.3. Utilitarian Social Welfare with Individual Utility Separable in the Distribution of Opportunity

Further increasing concreteness, I now assume that individual utility functions are additively separable in the distribution of opportunity alone, individuals not being concerned with the distribution of chosen actions and realized individual utility. Thus, individual utility has the form

$$u_j(z_\varphi) = v_j(c_{\varphi j}) + w_j[P(x, C_\varphi)], \tag{3}$$

where $w_j(\cdot)$ maps the distribution of covariates and choice sets into the real line.

Utility function (3) is relatively simple, yet it is general enough to express a wide spectrum of personal preferences. The functions $v_j(\cdot)$ and $w_j(\cdot)$ express the extent to which j values his own chosen action relative to the population distribution of opportunity. The form of $w_j(\cdot)$ expresses how j values population equity.

If individuals have interpersonally comparable cardinal utility functions of form (3), utilitarian social welfare under policy $\varphi$ is

$$(4) \quad W[u_j(z_\varphi), j \in J] = E[v(c_\varphi)] + E\{w[P(x, C_\varphi)]\}.$$

The expectations are taken across the population. $E[v(c_\varphi)]$ is mean self-centered welfare and $E\{w[P(x, C_\varphi)]\}$ is the mean distributional preference of the population. A utilitarian planner chooses a policy that maximizes (4) across $\varphi \in \Phi$, jointly considering the self-centered and distributional components of mean preferences.

### 3.4. Illustrations of Optimal Utilitarian Policy: Offering a Choice Option to a Subset of a Population

Increasing concreteness yet further, I present two illustrations of optimal utilitarian policy when social welfare has the utilitarian form (4). In both illustrations, each member of a population chooses an action from a choice set determined by a planner. Every person's choice set contains a default option, labeled A. Another option, labeled B, is available in sufficient quantity for the planner to offer it to a maximum fraction $\varphi_{max}$ of the population. Individuals have distributional preferences, valuing breadth and equality of opportunity. Members of the population interact infinitesimally.

In Section 3.4.1, individuals are observationally identical and value complete equality of opportunity. In Section 3.4.2, the population is composed to two observable groups and individuals value intergroup equality of opportunity.

3.4.1. Preference for Complete Equality of Opportunity in a Population of Observationally Identical Persons

Let members of the population be observationally identical from the perspective of the planner. Given this, the planner randomly selects the subset of persons who are offered option B.[19] The planner may offer B to the maximal feasible fraction $\varphi_{max}$ of the population or to some smaller fraction. Thus, the feasible policies are fractions in the interval $\Phi = [0, \varphi_{max}]$.

Consider policy $\varphi \in \Phi$. Let each individual j have a known utility function of form (3). If j is not offered option B, the choice set is $C_{\varphi j} = \{A\}$ and the self-centered component of utility is $v_j(A)$. If j is offered B, the choice set is $C_{\varphi j} = \{A, B\}$ and self-centered utility for the better action is $\max[v_j(A), v_j(B)]$. Mean self-centered utility with rational choice behavior is

$$(5) \quad E[v(c_\varphi)] = E[v(A)](1 - \varphi) + E\{\max[v(A), v(B)]\}\varphi = E[v(A)] + E\{\max[v(A), v(B)] - E[v(A)]\}\varphi.$$

Given that $E\{\max[v(A), v(B)]\} \geq E[v(A)]$, the policy that maximizes mean self-centered utility is $\varphi = \varphi_{max}$.

Individual j has distributional preferences that value both the breadth and equality of opportunity. Assume that these preferences have the form

$$(6) \quad w_j[P(x, C_\varphi)] = \alpha_j\varphi - \beta_j\varphi(1 - \varphi).$$

Here $\alpha_j \geq 0$ and $\beta_j \geq 0$ are person-specific parameters expressing the extent to which j values increasing the breadth and equality of opportunity respectively. When $\alpha_j > 0$, $\alpha_j\varphi$ expresses a preference for greater breadth of opportunity: utility increases linearly with the fraction of the population who face the larger choice set {A, B}. When $\beta_j > 0$, the quadratic function $-\beta_j\varphi(1 - \varphi)$ expresses a preference for greater equality of opportunity. The function is maximized at the value zero when $\varphi$ equals zero or one, each of which gives complete equality of opportunity. It is minimized at the value $-\beta_j/4$ when $\varphi = ½$, where inequality of opportunity is greatest. With this utility specification, the mean distributional preference is

$$(7)\quad E\{w[P(x, C_\varphi)]\} = E(\alpha)\varphi - E(\beta)\varphi(1 - \varphi).$$

[19] Random selection implies ex ante equal treatment of equals. Equal treatment does not occur ex post except in the polar cases where no one or everyone is offered option B.

Utilitarian social welfare adds mean self-centered utility and distributional preferences, giving

$$(8)\quad W[u_j(z_\varphi), j \in J] = E[v(A)] + E\{\max[v(A), v(B)] - E[v(A)]\}\varphi + E(\alpha)\varphi - E(\beta)\varphi(1 - \varphi)$$

$$= E[v(A)] + \big[E\{\max[v(A), v(B)]\} - E[v(A)] + E(\alpha) - E(\beta)\big]\varphi + E(\beta)\varphi^2.$$

A planner who knows $E[v(A)]$, $E\{\max [v(A), v(B)]\}$, $E(\alpha)$, and $E(\beta)$ can compute social welfare for each feasible value of $\varphi$ and choose a policy that maximizes social welfare.

Given that $E\{\max[v(A), v(B)]\} - E[v(A)] \geq 0$, utilitarian welfare weakly increases with $\varphi$ if $E(\alpha) - E(\beta) \geq 0$. Hence, the optimal policy is $\varphi_{max}$ in this setting. If $E(\alpha) - E(\beta) < 0$, social welfare remains an increasing function of $\varphi$ and the optimal policy remains $\varphi_{max}$ if $E(\alpha) - E(\beta)$ is less negative than $E\{\max[v(A), v(B)]\} - E[v(A)]$ is positive.

The behavior of social welfare as a function of $\varphi$ is more subtle if $E(\alpha) - E(\beta)$ is more negative than $E\{\max[v(A), v(B)]\} - E[v(A)]$ is positive. Then the quadratic function (8) is minimized at a value of $\varphi$ in the interval [0, ½]. Differentiating (8) with respect to $\varphi$ yields the first order condition for minimum welfare, which occurs at the value

$$(9)\quad \varphi_{foc} = ½\big\{E(\beta) - E(\alpha) - [E\{\max[v(A), v(B)]\} - E[v(A)]]\big\}/E(\beta).$$

If $\varphi_{max} \leq \varphi_{foc}$, social welfare decreases with $\varphi$ throughout the interval $[0, \varphi_{max}]$. Hence, the optimal policy is $\varphi = 0$, yielding social welfare $E[v(A)]$. In this setting, the mean preference for equality of opportunity is so strong that it is socially optimal to prohibit access to option B, even though making it available would increase mean self-centered utility.

If $\varphi_{max} > \varphi_{foc}$, social welfare decreases with φ in the interval $[0, \varphi_{foc}]$ and increases in the interval $[\varphi_{foc}, \varphi_{max}]$. The optimal policy is $\varphi = 0$ if (8) evaluated at $\varphi_{max}$ is less than E[v(A)]. It is $\varphi = \varphi_{max}$ if (8) evaluated at $\varphi_{max}$ is greater than E[v(A)]. Thus, optimal policy varies discontinuously with mean population preferences.

### 3.4.2. Preference for Intergroup Equality of Opportunity in a Population with Two Observable Groups

Again let option B be available in sufficient quantity to offer it to a maximum fraction $\varphi_{max}$ of the population. Again the planner may offer B to any fraction $\varphi \leq \varphi_{max}$.

In contrast to Section 3.4.1, let each member of the population have a binary observable covariate x taking the value 0 or 1. The planner may differentiate these groups when offering B, selecting recipients of the offer randomly within each group. Thus, the planner may select group-specific fractions $(\varphi_0, \varphi_1)$ to receive the offer, subject to the constraint $P(x = 0)\varphi_0 + P(x = 1)\varphi_1 \leq \varphi_{max}$.

*Optimizing Mean Self-Centered Utility*

As in Section 3.4.1, each individual has a known utility functions of form (3). If only A is available to individual j, self-centered utility is $v_j(A)$. If both A and B are available, self-centered utility for the better action is again $\max[v_j(A), v_j(B)]$. Hence, mean self-centered utility with rational choice behavior is

$$(10) \quad E[v(c_\varphi)] = \{E[v(A)|x = 0](1 - \varphi_0) + E\{\max[v(A), v(B)]|x = 0\}\varphi_0\}P(x = 0)$$
$$+ \{E[v(A)|x = 1](1 - \varphi_1) + E\{\max[v(A), v(B)]|x = 1\}\varphi_1\}P(x = 1)$$

$$= E[v(A)] + \{E\{\max[v(A), v(B)]|x = 0\} - E[v(A)|x = 0]\}P(x = 0)\varphi_0$$
$$+ \{E\{\max[v(A), v(B)]|x = 1\} - E[v(A)|x = 1]\}P(x = 1)\varphi_1.$$

Given that $E\{max[v(A), v(B)]|x = k\} \geq E[v(A)|x = k]$ for $k \in \{0, 1\}$, mean self-centered utility increases with both $\varphi_0$ and $\varphi_1$. This implies that maximization of mean self-centered utility occurs on the boundary of the feasibility constraint $P(x = 0)\varphi_0 + P(x = 1)\varphi_1 \leq \varphi_{max}$; thus, $\varphi_1$ should satisfy the equation $P(x = 1)\varphi_1 = \varphi_{max} - P(x = 0)\varphi_0$. Substituting this into (10) yields

$$(11)\quad E[v(c_\varphi)] = E[v(A)] + \Big[\{E\{max[v(A), v(B)]|x = 0\} - E[v(A)|x = 0]\} - \{E\{max[v(A), v(B)]|x = 1\} - E[v(A)|x = 1]\}\Big]P(x = 0)\varphi_0 + \{E\{max[v(A), v(B)]|x = 1\} - E[v(A)|x = 1]\}P(x = 1)\varphi_{max}.$$

Suppose that $E\{max[v(A), v(B)]|x = 0\} - E[v(A)|x = 0] > E\{max[v(A), v(B)]|x = 1\} - E[v(A)|x = 1]$; that is, offering B is more valuable, on average, to group 0 than to group 1. Then the right-hand side of (11) is increasing in $\varphi_0$. It follows that the policy that maximizes population-wide mean self-centered utility is $[\varphi_0 = \varphi_{max}/P(x = 0), \varphi_1 = 0]$ if $\varphi_{max} \leq P(x = 0)$ and is $[\varphi_0 = 1, \varphi_1 = [\varphi_{max} - P(x = 0)]/P(x = 1)]$ if $\varphi_{max} \geq P(x = 0)$. An analogous derivation holds if offering B is more valuable on average to group 1 than to group 0.

*Optimizing Utilitarian Welfare with Distributional Preferences*

The above derivation shows that maximization of mean self-centered utility may yield considerable intergroup inequality of opportunity. This may change if the population sufficiently values intergroup equality.

As in Section 3.4.1, assume that an individual may value both the breadth and equality of opportunity. In contrast to the earlier illustration, suppose that individuals value intergroup equality rather than complete equality. To formalize this, assume that distributional preference has the form

$$(12)\qquad w_j[P(x, C_\varphi)] = \gamma_j[P(x = 0)\varphi_0 + P(x = 1)\varphi_1] - \delta_j(\varphi_1 - \varphi_0)^2.$$

Here $\gamma_j \geq 0$ and $\delta_j \geq 0$ are person-specific parameters expressing the extent to which j values increasing the breadth and equality of opportunity respectively. When $\gamma_j > 0$, $\gamma_j[P(x = 0)\varphi_0 + P(x = 1)\varphi_1]$ expresses a preference for greater breadth of opportunity: utility increases linearly with the fraction of the population who face the larger choice set {A, B}, regardless of whether they are in group 0 or 1. When $\delta_j > 0$, the quadratic function $-\delta_j(\varphi_1 - \varphi_0)^2$ expresses a preference for greater intergroup equality of opportunity. The function is maximized at the value zero when $\varphi_1 = \varphi_0$. It is minimized when ($\varphi_0 = 0$, $\varphi_1 = \min[1, \varphi_{max}/P(x = 1)]$ or ($\varphi_0 = \min[1, \varphi_{max}/P(x = 0)]$, $\varphi_1 = 0$), where intergroup inequality of opportunity is greatest. With this utility specification, the mean distributional preference is

(13) $$E\{w[P(x, C_\varphi)]\} = E(\gamma)[P(x = 0)\varphi_0 + P(x = 1)\varphi_1] - E(\delta)(\varphi_1 - \varphi_0)^2,$$

Utilitarian social welfare adds mean self-centered utility and distributional preferences, giving

(14) $$\begin{aligned} W[u_j(z_\varphi), j \in J] = {} & E[v(A)] + [E\{\max[v(A), v(B)|x = 0]\} - E[v(A)|x = 0]]P(x = 0)\varphi_0 \\ & + [E\{\max[v(A), v(B)|x = 1]\} - E[v(A)|x = 1]]P(x = 1)\varphi_1 \\ & + E(\gamma)[P(x = 0)\varphi_0 + P(x = 1)\varphi_1] - E(\delta)(\varphi_1 - \varphi_0)^2 \end{aligned}$$

$$\begin{aligned} = {} & E[v(A)] + [E\{\max[v(A), v(B)|x = 0]\} - E[v(A)|x = 0] + E(\gamma)]P(x = 0)\varphi_0 \\ & + [E\{\max[v(A), v(B)|x = 1]\} - E[v(A)|x = 1] + E(\gamma)]P(x = 1)\varphi_1 \\ & - E(\delta)(\varphi_1 - \varphi_0)^2. \end{aligned}$$

A planner who knows $P(x)$, $\{E[v(A)|x = k], E\{\max[v(A), v(B)]|x = k\}, k = 0, 1\}$, $E(\gamma)$, and $E(\delta)$ can compute social welfare for each feasible value of $(\varphi_0, \varphi_1)$ and choose a policy that maximizes social welfare.

If one constrains policy choice to ones with intergroup equality (that is, $\varphi_1 = \varphi_0$), utilitarian welfare (14) reduces to (5), mean self-centered utility without observation of x. The optimal constrained policy is $\varphi_1 = \varphi_0 = \varphi_{max}$. In general, this is not the optimal unconstrained policy. The planner may be able to achieve greater social welfare by permitting $\varphi_1$ to differ from $\varphi_0$. Then society bears the welfare loss of some intergroup inequality in return for the ability to target option B towards the group which would benefit more from its availability.

### 3.5. Numerical Computation of Optimal Utilitarian Policy

The simplicity of the illustrations in Section 3.4 should not obscure the fact that determination of optimal utilitarian policy can be challenging. When algebraic analysis is infeasible, one may approach optimization computationally. Monte Carlo analysis can approximate utilitarian welfare for any specified policy. One may then maximize approximate welfare over a tractable finite set of policies.

Let social welfare have form (4). To perform Monte Carlo analysis, draw a random sample of size N from J, say $J_N$, with attributes $[v_j(\cdot), w_j(\cdot), x_j, C_{\varphi j};\ j \in J_N]$. Specify a policy $\varphi$. Assuming rational choice behavior, determine the chosen action $c_{\varphi j}$ for each member of sample $J_N$. Use the sample averages $E_N[v(c_\varphi)]$, $P_N(x, C_\varphi)$, and $E_N\{w[P_N(x, C_\varphi)]\}$ to approximate $E[v(c_\varphi)]$, $P(x, C_\varphi)$, and $E\{w[P(x, C_\varphi)]\}$ respectively. The Strong Law of Large Numbers implies that these averages consistently estimate (4) as $N \to \infty$.

If the policy space $\Phi$ is finite and not too large, one can perform Monte Carlo analysis for all policies and approximate the optimal policy by solving the problem $\max_{\varphi \in \Phi} E_N[v(c_\varphi)] + E_N\{w[P_N(x, C_\varphi)]\}$.

### 3.6. Learning Population Preferences

Apart from the challenge of computing optimal policies, the primary impediment to implementation of the ideas developed in this paper is empirical. Determination of optimal policy requires knowledge of the preferences of population members. This information is unavailable at present. How then to proceed?

In the near term, one may perform sensitivity analysis. Recall that Samuelson (1947, p. 220) stated: "It is a legitimate exercise of economic analysis to examine the consequences of various value judgments, whether or not they are shared by the theorist." Lacking empirical knowledge, welfare-economic research can conjecture alternative distributions of population preferences and determine optimal policy under each conjecture.

In the near term, one may also study planning under uncertainty. I have performed such studies in multiple settings, with focus on evaluation of policy by maximum regret. Manski (2024) provides a comprehensive exposition. The applications to date have addressed utilitarian planning assuming that individuals are self-centered. In principle, study of planning under uncertainty can examine settings where individuals have distributional preferences whose properties are not completely known.

The long term goal should be to perform new empirical research that yields knowledge of population preferences. I do not think it realistic to expect much progress based on econometric analysis of revealed preferences. The feasibility of credible revealed preference analysis is limited even when one finds it realistic to assume that preferences are self-centered; see, for example, the discussion of revealed preference analysis of (consumption, leisure) preferences in Manski (2014). Revealed preference analysis seeking to learn distributional preferences is yet more challenging.

I think it more promising to focus future research on performance of stated-choice analyses, in which a researcher poses multiple hypothetical choice settings to survey respondents and ask them to predict the choices they would make in each setting. If the respondents are a random sample of the population of interest, the data may be used to estimate the distribution of preferences in the population. Inference on preferences from data on stated choices has a long history in econometric analysis of discrete choice. Ben Akiva, McFadden, and Train (2019) provide a comprehensive review of the conventional practice, which asks persons to state deterministic choices. Concerned that researchers usually pose incomplete scenarios that do not fully describe choice settings, Manski (1999) proposed elicitation of choice probabilities.[20]

[20] Stated-choice analysis should not be confused with the qualitative attitudinal research long performed by sociologists and social psychologists. An example of attitudinal research in economics is Stantcheva (2021), who reported findings on perceptions of the fairness of income and estate taxes elicited from survey respondents. The question posed asked respondents if they perceive aspects of the current tax structure to be "fair" or "unfair," without defining these words.

Whatever the specifics of data collection and interpretation, a huge advantage of stated-choice analysis over revealed-preference analysis is that the choice settings considered are not limited by what nature offers. A researcher can elicit predictions of behavior in a wide spectrum of hypothetical choice settings.

Stated-choice data have been widely used in utilitarian assessment of medical treatments, under the assumption that preferences are self-centered. The prevalent practice is to measure health-related cardinal utility on a scale called a quality-adjusted life year (QALY). Weinstein, Torrance, and McGuire (2009) describe the scale as follows (p. S5):

> "Health states must be valued on a scale where the value of being dead must be 0, because the absence of life is considered to be worth 0 QALYs. By convention, the upper end of the scale is defined as perfect health, with a value of 1. To permit aggregation of QALY changes, the value scale should have interval scale properties such that, for example, a gain from 0.2 to 0.4 is equally valuable as a gain from 0.6 to 0.8."

Although the basic QALY scale is [0, 1], health economists often consider the possibility that persons may view some adverse health states as having value less than 0, thus extending the scale.

Although QALY measurement has assumed that individual preferences are self-centered, health economists have on occasion used stated-choice experiments to study preferences for various types of health equity. See, for example, Dolan (1998), Cookson *et al*. (2021), and Robson *et al*. (2024).

## 4. Conclusion

No external social planner exists in a democracy. Policy choices are made by a political process that need not maximize any social welfare function. In the United States, this process was established in the Constitution, which itself was created by a political process.

Researchers cannot definitively interpret what the framers of the Constitution had in mind when they wrote of the *general Welfare*. Nevertheless, welfare economics can contribute usefully to policy choice in democracies. Well-reasoned policy analysis is infeasible if the *general Welfare* remains an undefined concept. Study of policy choice with specified SWFs enables coherent analysis.

To be useful, I think it essential for welfare economics to move beyond conjecture of homogeneous, self-centered personal preferences. It should express the richness and variety of actual preferences over social states. I verbally distinguished various concepts of equity in Section 2. I formalized and illustrated a class of individual utility functions expressing distributional preferences in Section 3.

To close, I reiterate several themes that hold for utility functions with these types of distributional preferences. In summary:

(1) In a heterogeneous population, it is generally impossible to jointly achieve equality of opportunity and chosen actions.

(2) If preferences are monotonically separable and population interactions are infinitesimal, rational individual choice behavior under a specified policy is self-centered regardless of distributional preferences (Proposition 1). Nevertheless, if empowered to be the social planner, an individual with distributional preferences may choose to sacrifice some personal prosperity in order to enhance equity.

(3) Intergroup equality of opportunity is consistent with any degree of intragroup variation in choice sets (Proposition 2).